\documentclass[aps,prb,twocolumn,showpacs]{revtex4}
\usepackage{amsmath}
\usepackage{subfigure}
\usepackage{graphicx}
\usepackage{epsfig}
\usepackage{txfonts}
\usepackage{epsf}

\newlength{\upit}\upit=0.1truein

\newcommand{\ltappr}{{{\lower4pt\hbox{$<$} } \atop \widetilde{ \ \ \ }}}
\newlength{\bxwidth}\bxwidth=1.5 truein

\newlength{\figwidth}
\newlength{\shift}
\usepackage{epsf}

\newcommand \bea {\begin{eqnarray} }
\newcommand \eea {\end{eqnarray}}

\newcommand{\beg}{\begin{equation}}
\newcommand{\en}{\end{equation}}

\newcommand{\dg}{^{\dagger }}
\newcommand{\bk}{\mathbf k}
\newcommand{\bq}{\mathbf q}

\begin{document}

\title{Symplectic large-$N$ theory of topological heavy-fermion semiconductors}
\author{Maxim Dzero}
\affiliation{Department of Physics, Kent State University, Kent, OH 44240, USA}

\date{\today}

\begin{abstract} 
I present a theory of topological heavy-fermion semiconductors based on the large-$N$ symplectic representation for the electron spin. 
The theory is exact in the limit when the number of spin flavors $N=2k$ is infinite. I find that both weak and strong topological insulating phases exist for $k<3$. Furthermore, for $k\geq 3$ the weak topological insulating state fully suppressed while only strong topological and trivial insulator states survive. In addition, using the mean-field theory results, I consider the tunneling into topologically trivial and non-trivial phases of a generic heavy-fermion insulators by calculating the differential tunneling conductance. The implications of the presented results for the existing heavy-fermion semiconductors
are discussed. 
\end{abstract}


\maketitle

\section{Introduction}
Heavy fermions are a class of complex materials in which the effective mass of conduction electrons greatly exceeds the bare electron mass due to the strong hybridization between the conduction and localized $f$-electrons \cite{ColemanReview1,ColemanReview2}. The properties of heavy-fermion materials present an interest for researchers from the perspective of both technology and fundamental science. While the technology-driven research on heavy fermions is mostly focused on problems related to energy conservation and storage, fundamental physics research takes  two interconnected directions: the investigation of the microscopic origins of unconventional superconductivity (with CeCu$_2$Si$_2$, UBe$_{13}$ and UPt$_3$ as first examples of this phenomenon in solid state system \cite{hfsc1,hfsc2,hfsc3}), and discovery and analysis of the novel states of quantum matter.

Despite many years of experimental and theoretical research, the microscopic mechanisms responsible for the emergence of various quantum states in these materials remain unclear. The hidden-order phase in URu$_2$Si$_2$ \cite{HOexp1,HOexp2,HOexp3, HOth1,HOth2,HOth3,HOth4,HOth5}, magnetic field-induced non-Fermi-liquid behavior in YbRh$_2$Si$_2$ \cite{YRS}, low-temperature metallic conductivity in heavy-fermion semiconductor SmB$_6$ and Ce-based compounds \cite{SmB6,KIReviews1,KIReviews2,KIReviews3}, superconducting response to Yb-doping in CeCoIn$_5$ \cite{Maple2011,Fisk2011,Cigdem2010}, La- and Y-doping in CeCu$_2$Si$_2$ \cite{Steglich1983,Ott1987} and recently discovered quantum criticality in $\beta$-YbAlB$_4$ \cite{Satoru} are just a few examples. All these states emerge via a physical process unique to heavy-fermion materials in which strong interactions between conduction and localized $f$-electrons operate in an environment of very strong spin-orbit coupling.

SmB$_6$ is a prototypical example of a heavy-fermion semiconductor \cite{SmB6,KIReviews1,KIReviews2,KIReviews3}. Interaction 
between Sm $p$- and $d$-orbital conduction electrons and localized $f$-electron states leads to an opening of the 
hybridization gap at $T^*\simeq 100$K. The average electron $f$-level occupancy is well below one, $n_f<1$, demonstrating the strongly mixed valent nature of this material \cite{SmB6mv1,SmB6mv2}. Transport measurements in SmB$_6$ show an increase 
in resistivity below $T\simeq 50$K and then saturates at very low (below 5K) temperatures \cite{SmB6mv2,SmB6resist1,SmB6resist2,SmB6resist3,SmB6resist4,SmB6resist5,Kim}.
The value of residual resistivity grows when the quality of the sample increases and is of the order of $\rho_{sat}\sim 30 ~\Omega\cdot$cm.
This value is incompatible with the one originating from the metallic conduction in the presence of disorder induced scattering. The origin of
low-temperature conductivity still remains poorly understood \cite{KIReviews3,SmB6resist1}, however, it was recently proposed\cite{TKI,TKILong} that SmB$_6$ is a topological insulator and finite low-temperature conductivity can be due to topologically protected metallic surface states at sample boundaries.  
 
There are two major theoretical challenges in understanding the anomalous transport properties of SmB$_6$. The first challenge is due
to the strong Hubbard interaction between the conduction and $f$-electrons. The second challenge is that the symmetry of the lowest lying crystalline field multiplets, which are hybridized with conduction electrons and determine the symmetry of the hybridization gap, is not known. These challenges make the formulation of full microscopic transport theory in SmB$_6$ quite challending. Nevertheless, some general features of the heavy-fermion semiconductors in regards to the topological features in their band structure can be described on a more general (i.e. less material dependent) level. One important question, for example, is the question of stability of weak and strong topological insulating phases depending on degeneracy of the local $f$-level.  
 
The basic model which is thought to capture the main aspects of the physics of the heavy-fermion semiconductors is the Anderson lattice model. The Hamiltonian can be written as a sum of the following three terms:
\beg\label{H}
H=H_c+H_f+H_h.
\en
Here the first term $H_c$ describes the conduction electrons
\beg\label{HKL}
\begin{split}
H_c=\sum_{\bk , \sigma }\xi_{\bk }c^{\dagger}_{\bk \sigma }c_{\bk \sigma }, \quad \xi_\bk=-\frac{t}{6}\sum\limits_{i=x,y,z}\cos k_i-\mu_c
\end{split}
\en
where $\sigma$ denotes electron's spin projection, $t$ is the hopping amplitude (equal to bandwidth) and $\mu_c$ is the chemical potential. 
Consequently, $f$-electrons are described by
\beg\label{Hf}
H_f=\sum\limits_{j}\sum\limits_{\alpha=\pm 1}\varepsilon_{f}^{(0)} f_{j\alpha}\dg f_{j\alpha}+{U}\sum\limits_{i\alpha}
f_{i\alpha}\dg f_{i\alpha}f_{i\overline{\alpha}}\dg f_{i\overline{\alpha}}+\sum\limits_{\langle ij\rangle}\sum\limits_{\alpha=\pm 1}t_{ij}^{(h)} f_{i\alpha}\dg f_{j\alpha}
\en
where $f_{j\alpha}\dg$ creates an $f$-electron on site $j$ in a state 
$\alpha$ of a lowest lying multiplet $N_\Gamma$-degenerate multiplet denoted by $\Gamma$ (below we consider Kramers doublet only, so $N_\Gamma=2$), $\varepsilon_{f}^{(0)}$ is the $f$-electron energy and $U>0$  is the Hubbard interaction between the $f$-electrons. The last term in (\ref{Hf}) yields a very weak hole-like dispersion for $f$-electrons to enforce the fully gapped insulating state. We emphasize that index $\alpha$ is not a spin index due to the presence of the strong spin-orbit coupling. 
Lastly, the third term in Eq. (\ref{H}) accounts for the interaction between conduction and the $f$ electrons
\beg
\begin{split}
&H_h=\sum\limits_{j,\alpha=\pm 1}\left[V_{i\sigma,j\alpha} {c}_{i\sigma}^{\dagger} {f}_{j\alpha}+ {\rm h.c.}\right], \\
&V _{ i\sigma,j\alpha}=V\sum_{\bk}[\Phi_{\Gamma\bk}]_{\alpha \sigma }
e^{i \bk \cdot({\bf R}_i-{\bf R}_j)}
\end{split}
\en

In Refs. [\onlinecite{TKI,TKILong}] the phenomenological analysis of the Anderson lattice model based on the low-energy expansion of the $f$-electron self-energy \cite{Phenom1,Phenom2} has been used to analyze the topological structure of the resulting heavy-fermion semiconducting state. In this paper, I will resort to a microscopic approach based on the large-$N$ slave-boson theory and analyze the topological structure of the insulating state.
I employ the symplectic SP($N$) ($N=2k, k=1,2,...$) representation for the electronic operators to properly describe time-reversal symmetry of the electronic states. In agreement with the previous results \cite{TKI,TKILong} I find that for $N=2$ and $N=4$ there appears two (weak and strong) topologically non-trivial states depending on the relative position between the renormalized $f$-level and the chemical potential of the conduction band. Moreover, I found that for the large value of $N>4$ there is only strong topological insulating state. In addition, I will discuss the tunneling into a topologically non-trivial heavy-fermion semiconductors. 

In the next Section I will present the large-$N$ mean-field theory of topological heavy-fermion semiconductors. 
and the calculation of the tunneling conductance of the surface states in weak topological insulator. 
The Section II is devoted to the discussion of the results and conclusions. 

\section{Symplectic slave-boson theory}
Large-$N$ slave-boson mean-field theories ($N$ is the degeneracy of the $f$-electron level) utilize the naturally small
parameter $1/N$ to determine the thermodynamic properties of heavy-fermion materials by expanding near exactly solvable limit
\cite{Barnes,SlaveBoson1,SlaveBoson2,kondolattice,SlaveBoson3,MillisLee,Hewson,Barzykin2006}
of $N\to\infty$. Recently, a novel large-$N$ expansion methods have been developed to account for the specific symmetries of the problem (see [\onlinecite{FlintNature}] and references therein). 
In what follows, we generalize our model from SU($2$) symmetry group to SP($N$) with $N=2k$, so that the spin summation run over $k$ spin indices, $\alpha,\sigma\in[\pm 1,\pm k]$. The importance of using the SP($N$) subset of SU($N$) group clearly lies in the requirement for the
proper description of the states related by time-reversal \cite{FlintNature}. In the SP($N$) version of the theory, 
the form-factor matrix acquires a block-diagonal form of identical $2\times 2$ blocks. For the subsequent saddle-point analysis 
we find it more convenient to use the path integral formulation. 

The slave-boson approximation corresponds to (i) taking the limit $U\to\infty$, which corresponds to projecting out the doubly occupied states and (ii) introducing the constraint
which guarantees the local moment at the $f$-site, i.e. $n_f=1$: 
\beg\label{constraint}
\begin{split}
& U\to\infty: \quad f_{i\alpha}\to f_{i\alpha}b_i\dg, \quad f_{i\alpha}\dg\to f_{i\alpha}\dg b_i,\\
& \sum\limits_{\alpha}f_{i\alpha}\dg f_{i\alpha}+
b_i\dg b_i=1.
\end{split}
\en

The partition function corresponding to the model Hamiltonian (\ref{H}) with constraint condition (\ref{constraint}) reads :
\beg
Z=\int\limits_{-\pi/\beta}^{\pi/\beta}\frac{\beta d\lambda}{\pi}\int{\cal D}(b,b\dg,f,f\dg,c,c\dg)
\exp\left(-\int\limits_0^\beta L(\tau)d\tau\right), 
\en
where Lagrangian $L(\tau)$ is 
\beg
\begin{split}
L=&\sum\limits_{i}b_i\dg\frac{d}{d\tau}b_i+\sum\limits_{ij}\sum\limits_{\alpha=1}^{N}f_{i\alpha}\dg\left[\delta_{ij}\left(\frac{d}{d\tau}+\varepsilon_f^{(0)}\right)+b_i t_{ij}^{(h)}b_j\dg\right)f_{j\alpha}\\&+\sum\limits_{\bk}\sum\limits_{\alpha=1}^{N}c_{\bk\alpha}\dg\left(\frac{d}{d\tau}+\xi_\bk\right)c_{\bk\alpha}\\&
+\frac{V}{\sqrt{N}}\sum\limits_{i,\bk}\sum\limits_{\alpha,\beta=1}^N\left([\Phi_{\Gamma\bk}]_{\alpha\beta}
e^{i \bk \cdot{\bf R}_i}f_{i\alpha}\dg b_{i}c_{\bk\beta}+\textrm{h.c.}\right)\\&+\sum\limits_{j}i\lambda_j\left(\sum\limits_{\alpha=1}^Nf_{j\alpha}\dg f_{j\alpha}+b_j\dg b_j-1\right)
\end{split}
\en
Here we use the same notation for the form-factor matrix, although now it has a block diagonal form of $k(=N/2)$ blocks each of dimension $2\times 2$. We have also introduced the field $\lambda_j$ to enforce a constraint. 
Finally, we have rescaled the hybridization amplitude $V\to V/\sqrt{N}$ for the bookkeeping purposes (see below).

Now we can integrate the conduction electrons by making the following transformation
\beg
\begin{split}
&c_{\bk\alpha}\to c_{\bk\alpha}-\frac{V}{\sqrt{N}}\sum\limits_{i}\sum\limits_{\beta=1}^N[\Phi_{\Gamma\bk}^*]_{\alpha\beta}
e^{-i \bk \cdot{\bf R}_i}(\partial_\tau+\xi_\bk)^{-1}f_{i\beta} b_{i}\dg, \\
&c_{\bk\alpha}\dg\to c_{\bk\alpha}\dg-\frac{V}{\sqrt{N}}\sum\limits_{i}\sum\limits_{\beta=1}^N[\Phi_{\Gamma\bk}]_{\beta\alpha}
e^{i \bk \cdot{\bf R}_i}f_{i\beta}\dg b_{i}(\partial_\tau+\xi_\bk)^{-1}
\end{split}
\en
In what follows, it is convenient to write the $2\times 2$ form-factor matrix as follows
\beg
\Phi_{\Gamma \bk}=\phi_\bk({\vec n}_\bk\cdot{\vec \tau}), \quad \phi_\bk^2={\frac{1}{2}\textrm{Tr}[\Phi_{\Gamma\bk}\dg\Phi_{\Gamma\bk}]}
\en
where ${\vec \tau}$ are Pauli matrices and ${\vec n}_\bk$ is a unit vector. 
We obtain
\beg
\begin{split}
&L=\sum\limits_{i}b_i\dg\frac{d}{d\tau}b_i+\sum\limits_{ij}\sum\limits_{\alpha=1}^{N}f_{i\alpha}\dg\left[\delta_{ij}\left(\frac{d}{d\tau}+\varepsilon_f^{(0)}\right)+b_i t_{ij}^{(h)}b_j\dg\right]f_{j\alpha}\\+&
\sum\limits_{\bk}\sum\limits_{\alpha=1}^{N}c_{\bk\alpha}\dg\left(\frac{d}{d\tau}+\xi_\bk\right)c_{\bk\alpha}
+\sum\limits_{j}i\lambda_j\left(\sum\limits_{\alpha=1}^Nf_{j\alpha}\dg f_{j\alpha}+b_j\dg b_j-1\right)
\\&-\frac{|V|^2}{N}\sum\limits_{ij,\bk}\sum\limits_{\alpha,\beta=1}^N\Delta_{\alpha\beta}(\bk)
e^{i \bk \cdot({\bf R}_i-{\bf R}_j)}f_{i\alpha}\dg b_{i}(\partial_\tau+\xi_\bk)^{-1}f_{j\beta} b_{j}\dg, 
\end{split}
\en
where we have introduced
\[
\Delta_{\alpha\beta}(\bk)=\frac{1}{N}\sum\limits_{\gamma=1}^N[\Phi_{\Gamma\bk}^*]_{\alpha\gamma}
[\Phi_{\Gamma\bk}]_{\gamma\beta}=\phi_\bk^2\delta_{\alpha\beta}
\]

The action with the Lagrangian above is quadratic in fermionic operators, which can be integrated out to give an effective action
in terms of the slave fields only. Since the resulting expression is quite cumbersome we will not give it here. 
Instead, we proceed with the saddle-point analysis. 
%%%%%%%%%%%%%%%%% Fig. 1 - Diagram %%%%%%%%%%%%%%%%%
\begin{figure}[h]
\includegraphics[width=2.4in,angle=0]{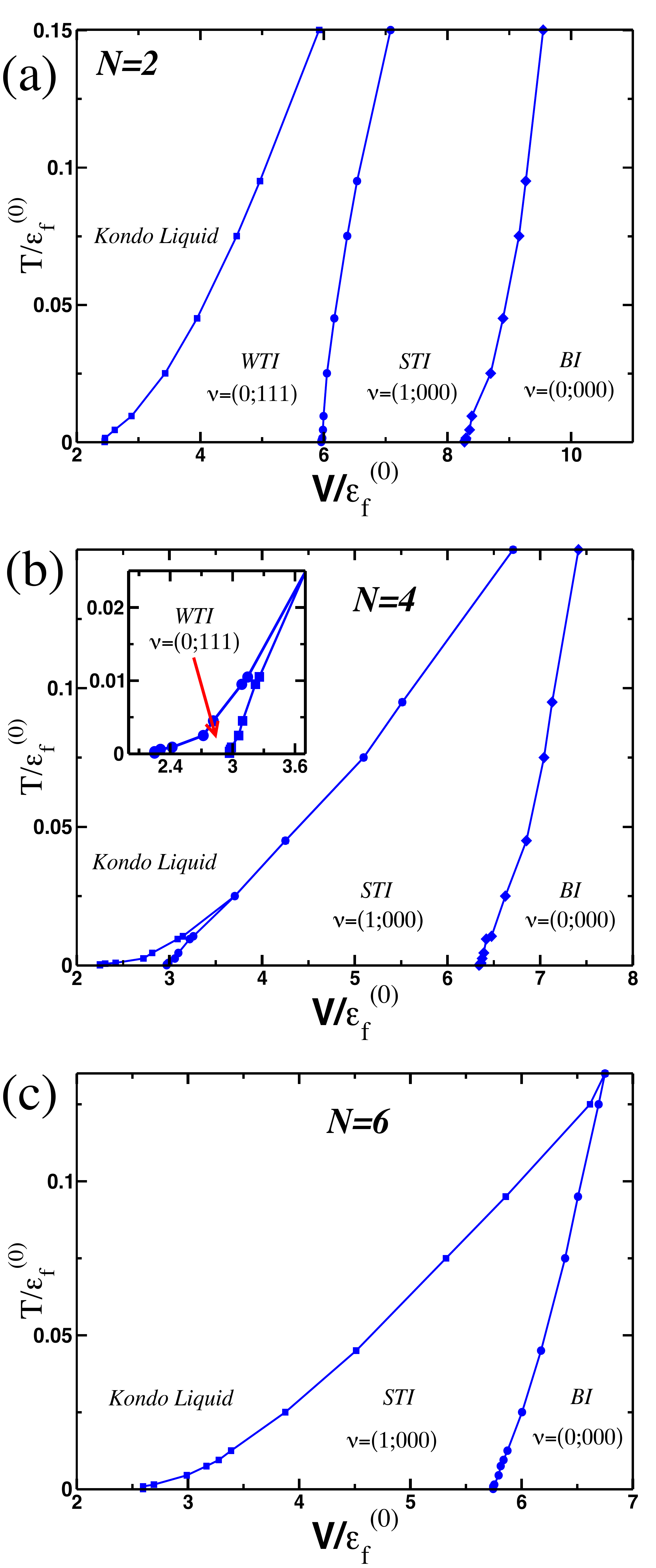}
\caption{Phase diagram found from the solution of the slave-boson mean-field equations (\ref{MFEqs}). Kondo liquid state corresponds to 
the situation when the slave-boson amplitude $a=0$. I have used the following
values for the input parameters: $t_f^{(h)}=0.1t$, $\varepsilon_f^{(0)}=-1.05t$. When the number of the fermionic flavors exceeds four, 
$N>4$, there exists only strong topological insulating phase and weak topological insulating state disappears.}
\label{Fig1TKI}
\end{figure}
%%%%%%%%%%%%%%%%% End of Fig. 1 - Diagram %%%%%%%%%%%%%%%%%

\subsection{mean-field solution}
Mean-field (saddle-point) approximation corresponds to the following values of the bosonic fields:
\beg
b_\bq(\tau)=\sqrt{N}a\delta_{q,0}, \quad i\lambda_\bq(\tau)=(\varepsilon_f-\varepsilon_{f}^{(0)})\delta_{q,0},
\en
where both $a$ and $\varepsilon_f$ are $\tau$-independent. Now, we can use the Matsubara frequency representation and 
integrate out $f$-fields, since the action is quadratic in these fields. These yields:
\beg
\begin{split}
Z=&\int\limits_{-\pi/\beta}^{\pi/\beta}\frac{\beta d\lambda}{\pi}\int{\cal D}b{\cal D}b\dg e^{-S_{eff}}, \\
S_{eff}=&N\left(\varepsilon_f-\varepsilon_f^{(0)}\right)(a^2-q_N)\\&-2NT\sum\limits_{i\omega}\sum\limits_{\bk}
\log[(i\omega-\omega_{1\bk})(i\omega-\omega_{2\bk})], 
\end{split}
\en
where we have introduce the parameter $q_N=\frac{1}{N}$. Moreover, functions $\omega_{1,2\bk}$ describe newly formed energy 
bands
\beg\label{hfbands}
\begin{split}
\omega_{1,2\bk}&=\frac{1}{2}\left[\xi_\bk+E_{f\bk}\pm\sqrt{(\xi_\bk-E_{f\bk})^2+4(Va\phi_\bk)^2}\right], \\
E_{f\bk}&=\varepsilon_f+Na^2h_\bk, \quad h_\bk=\frac{1}{6}t_f^{(h)}\sum\limits_{i=x,y,z}\cos k_i.
\end{split}
\en
We note that the newly formed band spectrum corresponds to the effective Hamiltonian
\beg\label{Heff}
\mathcal{H}_{eff}(\bk)=
\left(
\begin{matrix}
\xi_\bk\underline{1} & {Va}{\Phi}_{\Gamma\bk}\dg \\
{Va}{\Phi}_{\Gamma\bk} & E_{f\bk}\underline{1}
\end{matrix}
\right),
\en
The reason we invoke the effective Hamiltonian is that it will allow us to analyze the topological structure of an insulating state. 
In Eq. (\ref{Heff}) $\underline{1}$ denotes the unit $2\times2$ matrix. To determine the parameters $a$ and $\varepsilon_f$ we, of course,  have to minimize the effective action. In addition, we have to keep in mind that
the total number of electrons must be conserved. Specifically, for an insulator, we have to require that we will have one conduction electron per $f$-electron, so that
\beg\label{insulator}
n_c+n_f=N.
\en

Minimization of the effective action together with the condition for an insulator (\ref{insulator}) yields the following system of the mean-field
equations:
\beg\label{MFEqs}
\begin{split}
&(\varepsilon_f-\varepsilon_f^{(0)})a+T\sum\limits_{i\omega,\bk}\left[Na h_\bk A_{ff}(\bk,i\omega)+V\phi_\bk A_{fc}(\bk,i\omega)\right]=0, \\
&(a^2-q_N)+T\sum\limits_{i\omega}\sum\limits_\bk A_{ff}(\bk,i\omega)=0,\\
&(q_N-a^2)+T\sum\limits_{i\omega}\sum\limits_\bk A_{cc}(\bk,i\omega)=1,
\end{split}
\en
where the functions $A_{ab}(\bk,i\omega)$ are defined by 
\beg
\begin{split}
&A_{ff}(\bk,i\omega)=\frac{i\omega-\xi_\bk}{(i\omega-\xi_\bk)(i\omega-E_{f\bk})-V^2a^2\phi_\bk^2}, \\
&A_{fc}(\bk,i\omega)=\frac{Va\phi_\bk}{(i\omega-\xi_\bk)(i\omega-E_{f\bk})-V^2a^2\phi_\bk^2}, \\
&A_{cc}(\bk,i\omega)=\frac{i\omega-\xi_\bk}{(i\omega-E_{f\bk})(i\omega-E_{f\bk})-V^2a^2\phi_\bk^2}. \\
\end{split}
\en
To solve (\ref{MFEqs}) we still need to specify the momentum dependence of the hybridization gap, $\phi_\bk$. 
In what follows, we adopt the choice of the form-factors from Refs. [\onlinecite{TKI,TKILong}] and consider function 
$\phi_\bk$ which at small momenta $\bk$
is  $\phi_{\hat\bk}=\frac{1}{12}\sqrt{\frac{3}{\pi}}[12\cos(2\theta)+5(3+\cos(4\theta))]^{1/2}$,
where $\theta$ define the direction of the unit vector
$\hat{\bf k}$, associated with the point on the Fermi surface. 

To analyze the topology of the bands governed by the effective Hamiltonian (\ref{Heff}) we  use the fact that 
topology  is invariant under any adiabatic deformation
of the Hamiltonian. We begin our study with a 
tight-binding model for a KI on a simple cubic lattice. Our choice of hybridization ensures that 
the mean-field Hamiltonian (Eq. \ref{Heff}) is
a periodic function satisfying $\mathcal{H}_{eff}({\bf
k})=\mathcal{H}_{eff}({\bf k}+{\bf G})$.
The technical analysis is readily generalized to 
more complicated cases as discussed below.  The most important element 
of the analysis is the odd parity form factor of the 
$f$ electrons, ${\Phi}_{\Gamma}(\bk)=-{\Phi}_{\Gamma}(-\bk)$.  
This parity property together with the absence of the nodes in the hybridization gap across the Brillouin zone (BZ)
are the only essential input as far as the topological
structure is concerned.

To evaluate the topological indices we use the results of 
Fu and Kane [\onlinecite{FuKane2007}] who have demonstrated that in an insulator
{\em with time-reversal and space-inversion symmetry}, the topological
structure is determined by parity properties at the eight
high-symmetry points, $\bk^*_m$, in the 3D BZ, which are invariant
under time-reversal, up to a reciprocal lattice vector:
$\bk^*_m=-\bk^*_m+{\bf G}$.  In our
case, these symmetries require that $\mathcal{H}_{eff}({\bf k})={P}
\mathcal{H}_{eff}(-{\bf k}){P}^{-1}$ and $\mathcal{H}_{eff}({\bf
k})^{T}={\cal T} \mathcal{H}_{eff}(-{\bf k}){\cal T}^{-1}$, where
the parity matrix $P$
and the unitary part of the time-reversal
operator ${\cal T}$ are given by
\begin{equation}\label{}
P = \begin{pmatrix} \underline{1}& \cr & -\underline{1}\end{pmatrix},
\qquad 
{\cal T} = \begin{pmatrix}  i \sigma_{2}&\cr & i \sigma_{2}  \end{pmatrix},
\end{equation}
where $\sigma_{2}$ is the second Pauli matrix. 
For any space-inversion-odd form factor, it follows immediately that
$\hat{\Phi}_{\Gamma}(\bk)=0$ at a high-symmetry point. Hence, the
Hamiltonian at this high symmetry point is simply
$\mathcal{H}_{eff}({\bk^*_m})=(\xi_{\bk_m^*}+E_{f\bk_m^*})
I/2+(\xi_{\bk^*_m}-E_{f\bk_m^*}){P}/2$, where $I$ is 
the four-dimensional identity matrix.

The parity at a high symmetry point is thus determined by $\delta_m=\textrm{sgn}(\xi_{\bk^*_m}-E_{f\bk_m^*})$.
Four independent $Z_2$ topological indices $(\nu_0;\nu_1,\nu_2,\nu_3)$ ~\cite{Kitaev}, one strong ($\nu_0$) and three weak indices 
($\nu_{1,2,3}$) can be constructed from $\delta_m$: (i)~The strong topological index is the product of all eight 
$\delta_m$'s: $I_{\rm STI} = (-1)^{\nu_0}=\prod\limits_{m=1}^{8} \delta_m = \pm 1$; 
(ii)~by setting $k_j=0$ (where $j= x,y, \mbox{and } z$),  
three high-symmetry planes, $P_j = \left\{ {\bf k}: k_j=0\right\}$, are formed that contain four high-symmetry points each. The product of the parities at these four points defines the
corresponding weak-topological index, $I_{\rm WTI}^{a}=(-1)^{\nu_\alpha}= \prod\limits_{{\bf k}_m \in P_j} \delta_m = \pm 1$ ($\alpha=1,2,3$) 
with integers corresponding to the axes $x,y$ and $z$. The existence of the three weak topological indices in 3D is related to a $Z_2$ topological index for 2D systems (a weak 3D TI is similar to a stack of 2D $Z_2$ topological insulators). 
Because there are three independent ways to stack 2D layers to form a 3D system,
the number of independent weak topological indices is also three.
A conventional band insulator has all of the four indices $I_{\rm STI}
= I_{\rm WTI}^x=I_{\rm WTI}^y=I_{\rm WTI}^z = +1$ or  equivalently (0;0,0,0). An index
$I=(-1)$ ($\nu_\alpha=1$) indicates a $Z_2$ topological state with the odd number of
surface Dirac modes. In a KI the symmetry index $\delta_{m}$ of a particular
high symmetry point $m$ is negative provided
$\xi_{\bk^*_{m}}<E_{f\bk_m^*}$ is lower the f-energy $E_{f\bk_m^*}$. 
Thus if 
$\xi_{{\bk_m^*}=0}<E_{f\bk_m^*}$  at the $\Gamma$ point, while
 $\xi_{\bf{{\bk^*_m\ne 0}}}>E_{f\bk_m^*}$ for all other symmetry
 points,  then $I_{\rm STI}
= -1$, and hence {the Kondo insulating state is a
strong-topological insulator, robust against disorder} \cite{TKI,TKILong}. Weak-topological insulators and topologically trivial
insulators can in principle be found for different band structures and
different values of $E_{f\bk_m^*}$. A particularly
interesting possibility is to tune topological phase transitions
between different types of insulators (e.g., by applying a pressure). 
Although we have been specifically considering a tight-binding
model with a primitive unit cell, all our conclusions apply directly
to systems adiabatically connected to this model. 

I solve the mean-field equations (\ref{MFEqs}) numerically. For a given value of hybridization and temperature I analyze the topological structure of the effective Hamiltonian (\ref{Heff}) using the prescription outlined above. The results are shown on Fig. 1. 
First I note that when $N=2$ the weak topological insulator (WTI) state precedes the strong topological insulating state in agreement 
with the earlier studies \cite{TKISU2}. For $N=4$ $(k=2)$ the region where WTI exists shrinks and is fully absent for $N=6$ $(k=3)$. It is a quite surprising observation for it is a special case when the slave-boson mean-field theory results crucially depend on the number of fermionic flavors. In other words, here we find an example when there is no adiabatic connection between the phases for the realistic case of $N=2$ and $N\to\infty$, i.e. when the mean-field theory is exact. The reason for the disappearance of WTI phase can be easily traced to the condition for the WTI: half of the $\delta_m$'s must be negative, while the other half must be positive. However, as it directly follows from the solution of the mean-field equations, this condition can never be satisfied when $k>2$. 
Lastly, the  results on Fig. 1 are consistent with the ones which have been obtained for the low-energy version of the Anderson model \cite{TKI,TKILong}. There it was found that WTI is stabilized for $f$-level energy close to the chemical potential for the conductions: this situation corresponds to the $f$-level occupation $n_f\simeq 1$. With an increase of hybridization, $n_f$ is reduced as system shows more mixed valent behavior, so that $n_f<1$. For $N=2$ as soon as insulator becomes a strong topological insulator, $n_f\simeq 0.8$. 

Generally, fluctuations around large-$N$ mean-field solution introduce interaction between the newly formed heavy-fermions.
Strictly speaking one needs to prove that fluctuations of the amplitude and phase of the slave-bosons do not break the newly formed state.  
Due to the presence of the bulk gap, however, we do not expect that fluctuations will lead to the substantial modification of the 
ground state. The separate issue, of course, is the effects of the fluctuations of the metallic surface states. Specifically, whether
the interactions may lead to an opening of gap at the surface. The detailed investigation of that problem goes beyond the scope
of this paper and we leave it for the future studies.

\subsection{tunneling into topological heavy-fermion semiconductors}
%
% Review the tunneling into f-electron systems. 
% Fano line shape for the tunneling conductance: interference between f- and c-states
% 
Scanning tunneling spectroscopy (STM) of heavy-fermion metals has become an active topic of experimental
and theoretical research in recent years \cite{STMexp1,STMexp2,STMexp3,STMtheory1, STMtheory2,STMtheory3}.
In this regard, an intriguing question is whether the STM can directly probe the metallic surface states
in a topological heavy-fermion semiconductor. In this Section I address this question by evaluating the 
differential tunneling conductance into a weak topological heavy-fermion semiconductor. In what follows I will use 
the mean-field theory discussed above for the case of SU(2) group, i.e. two flavors of fermions. 

To evaluate the tunneling conductance, we will model a bulk system by a stack of $L$ planes along the 
$z$-direction and diagonalize the Hamiltonian. The resulting Hamiltonian matrix $H_{nm}$ has blocks along the 
diagonal given by (\ref{Heff}), which describe the hopping and hybridization within each plane and the off-diagonal parts describing 
the hopping and hybridization between the planes. Since the in-plane momentum is a good quantum number, 
the dispersion of the conduction electrons is given by $\varepsilon(k_x,k_y)=-(t/4) (\cos k_x+\cos k_y)$, while the 
dispersion of the $f$ electrons is described by $\epsilon_f(k_x,k_y)=(t_f/4)(\cos k_x+\cos k_y)$ with $t_f=0.1t$. 
In addition, we have taken the form factor matrix in the form:
\begin{equation}\label{Phi}
\underline{\Phi } = \left\{
\begin{matrix}
V (\sin k_{x}\tau_{x}+ \sin k_{y}\tau_{y}), \textrm{ within the planes}, \\
iV_z\tau_z, ~\textrm{between the planes (upwards)}, \\
-iV_z\tau_z, ~\textrm{between the planes (downwards)}.
\end{matrix}
\right.
\end{equation}
Within each plane the conduction and $f$-electrons are described by the four-component spinor 
\beg\label{Spinor}
\Psi_{l\bk_\perp}\dg=(c_{l\bk_\perp,1}\dg ~c_{l\bk_\perp,2}\dg ~f_{l\bk_\perp,1}\dg ~f_{l\bk_\perp,2}\dg),
\en
where $l$ labels the layer and $\bk_\perp=(k_x,k_y)$. Lastly, the Hamiltonian describing
tunneling between an electrode and a sample is
\beg\label{Htun}
\begin{split}
H_{tun}&=T_c\sum\limits_{\bk_\perp,\sigma} (p_\sigma\dg c_{1\bk_\perp,\sigma}+\textrm{h.c.})\\&+T_f\sum\limits_{\sigma\beta,\bk_\perp}\left(
[\Phi_{\Gamma}(\bk_\perp)]_{\sigma\alpha}p_\sigma\dg f_{1\bk_\perp,\alpha}+\textrm{h.c.}\right)
\end{split}
\en
Here we have assumed that the tunneling involves conduction and $f$-electron states on the surface layer only ($l=1$). As we will see below, the presence of the form-factor in the second term is crucial for the cotunneling events, which ultimately give rise to the Fano lineshape for the differential tunneling conductance. 

%%%%%%%%%%%%%%%%% Fig. 2 - dI/dV %%%%%%%%%%%%%%%%%
\begin{figure}[h]
\includegraphics[width=2.5in,angle=0]{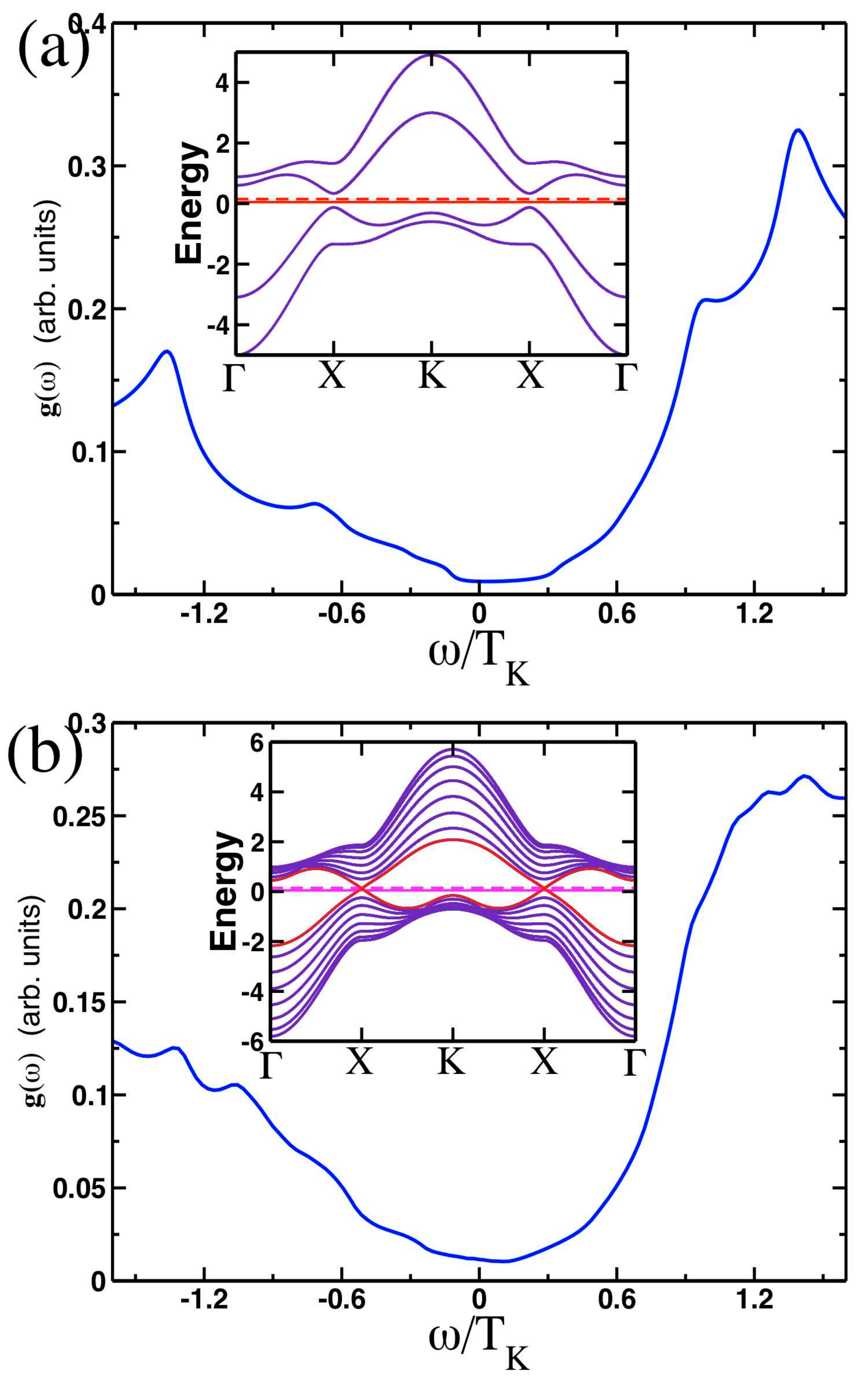}
\caption{Plots of the differential tunneling conductance $g(\omega)$ for the stack of $L$ layers of heavy-fermion semiconductors. 
Panel (a) shows $g(\omega)$ for the stack of $L=2$ layers: for this case there are no states in the gap. The band structure as a function of the momentum in the 2D BZ is shown on inset. Panel (b) shows $g(\omega)$ for the stack of $L=8$ layers corresponding to the weak topological insulator (even number of Dirac nodes in the gap). The asymmetry in the tunneling conductance is due to the cotunneling processes into conduction and $f$-electron states.}
\label{Fig2DOS}
\end{figure}
%%%%%%%%%%%%%%%%% End of Fig. 2 - dI/dV %%%%%%%%%%%%%%%%%

If we now assume that the tunneling electrode is in equilibrium state with the surface, the tunneling current as a function of the voltage, $I(V)$,
is 
\beg
\begin{split}
&I(V)=\frac{2e}{\hbar}\int\limits_{-\infty}^\infty d\omega \rho_{tip}(\omega-eV)[n_F(\omega-eV)-n_F(\omega)]\\ & 
\times\textrm{Im}\left[|T_c|^2G_{cc}(\omega)+|T_f|^2G_{ff}(\omega)+2|T_c||T_f|G_{cf}(\omega)\right],
\end{split}
\en
where $\rho_{tip}$ is the STM tip density of states (DOS), $n_F(\omega)$ is the Fermi distribution function and $G_{ab}(\omega)$
are the advanced local single particle Green functions:
\beg
G_{ab}(\omega)=\sum\limits_{\lambda,\bk_\perp}\frac{\phi_{a,\lambda}^*(\bk_\perp)\phi_{b,\lambda}(\bk_\perp)}{\omega-\epsilon_\lambda(\bk_\perp)-i\delta}.
\en
In the expression above, $\varepsilon_\lambda(\bk_\perp), \phi_{a,\lambda}(\bk_\perp)$ denotes the set of $\lambda$ eigenvalues and the corresponding eigenfunctions I obtained by diagonalizing the Hamiltonian $H_{nm}$, while subscripts $a,b$ refer to the components of the spinor (\ref{Spinor}) at the surface ($l=1$). In real materials, however, electronic correlations as well as disorder lead the broadening of the $f$-electron level. One way to account for these effects in the differential tunneling conductance $g(V)=dI/dV$ is to consider the complex
valued quasiparticle energies:
\beg
\epsilon_\lambda(\bk_\perp)\to E_{\lambda\bk_\perp}-i\Gamma_{\lambda\bk_\perp},
\en
with the quasiparticle width given by \cite{STMtheory3}
\beg\label{Gamma}
\Gamma_{\lambda\bk_\perp}=
\left\{
\begin{matrix}
\Gamma_{\lambda\bk_\perp}^2/T_K, \quad E_{\lambda\bk_\perp}<T^*, \\
|E_{\lambda\bk_\perp}|/[1+\log(|E_{\lambda\bk_\perp}|/T_K)]^2, \quad E_{\lambda\bk_\perp}>T^*.
\end{matrix}
\right.
\en
Here $T^*$ is the temperature corresponding to the opening of the hybridization gap, or the temperature at which the first non-trivial solution of the mean-field equations appears. In SmB$_6$, for example, $T^*\simeq 100$K.
With these provisions, we evaluate the differential tunneling conductance for the set of the parameters corresponding to the weak-topological 
insulator in 3D translationally invariant system. I show the results on Fig. 2 for a fully gapped states (top panel) and metallic state (bottom panel) corresponding to the weak topological insulator. Both curves have characteristic asymmetry due to the co-tunneling processes into conduction and $f$-states. The finite value of the $g(V)$ are zero bias are due to the finite width of quasiparticle states, Eq. (\ref{Gamma}). From our results we see that $g_{WTI}(V\sim 0)$ in the case of weak topological insulator is significantly enhanced in comparison with 
$g_{BI}(V\sim 0)$ (band insulator), which is not surprising. In addition, $g_{WTI}(V)$ shows higher asymmetry then $g_{BI}(V)$. Nevertheless, 
it is seems to be a quite challenging task to argue in favor of the topologically protected surface states solely on the STM data.

\section{Conclusions and outlook}
In this paper I have analyzed the low-temperature properties of the generic heavy-fermion semiconductors using the large-$N$ slave-boson theory. Specifically, I have provided an evidence for the formation of the topologically non-trivial electronic states at the surface of these materials. The results reported in this paper are in agreement with those obtained using different approach based on the low-energy analysis of the Anderson lattice model.  The phase diagram obtained within the mean-field analysis implies that the strong topological insulating phase is likely to be observed in materials with high, i.e. cubic,
point group symmetry. In this case, the analysis of the crystalline field split $f$-ion multiplets for the valence configurations corresponding to the total angular momentum $J=5/2$ or $J=7/2$ shows that only $N=4$ degenerate multiplets can give rise to the nodeless hybridization gap. Such a scenario can be realized in heavy-fermion semiconductor SmB$_6$. Indeed, finite
metallic conductivity below $T\simeq 5$K may serve as a signature for topologically protected metallic surface states. The fact that conductivity grows with the improvement of the quality of the samples qualitatively supports this idea. Indeed, recent theoretical works \cite{STIdis1,STIdis2} have explicitly demonstrated that the presence of relatively strong disorder on the surface of a strong topological insulator will substantially disrupt these states. The physical reason for such a behavior is that 
the impurity induced states propagate well below leading to the diffusive behavior in the surface transport. In that regard, 
when the strength of disorder potential is comparable to the bulk gap, the 2D Dirac theory description of the 3D topological 
insulators is not valid \cite{STIdis2}. 

An important issue for the subsequent study is a role of fluctuations around the mean-field solution. For the band insulator, one may argue 
that the fluctuations effects (generally of the order of $1/N$) do not lead to any significant changes providing only small corrections to the gap itself. Situation becomes drastically different when the metallic surface states are present. In particular, fluctuations in slave-boson fields
lead to the effective interactions between the conduction and $f$-electrons, which in principle may lead to the opening of the gap at the
surface as well. The detailed investigation of these effects goes beyond the scope of this paper and I leave it for the future. 

\section{acknowledgments}
I thank Piers Coleman and Pedro Schlottmann for useful comments. 
Author acknowledges the financial support by the Ohio Board of Regents Research Incentive Program grant OBR-RIP-220573.
This work was supported in part by the National Science Foundation under grant No. 1066293 and the hospitality of the Aspen Center for Physics (M.D.)

\end{document}